# Data-driven design of high-temperature superconductivity among ternary hydrides under pressure

Bowen Jiang,[1,2,3] Xiaoshan Luo,[1,2] Toshiaki Iitaka[4], Ying Sun,[1,2] Xin Zhong,[1,2,]* Jian Lv,[1,2,3] Yu Xie,[1,]* Yanming Ma,[1,2,3] Hanyu Liu[1,]*

[1]*Key Laboratory of Material Simulation Methods and Software of Ministry of Education, College of Physics, Jilin University, Changchun 130012, China*

[2]*State Key Laboratory of Superhard Materials, College of Physics, Jilin University, Changchun 130012, China*

[3]*International Center of Future Science, Jilin University, Changchun 130012, China*

[4]*Discrete Event Simulation Research Team, RIKEN Center for Computational Science, 2-1 Hirosawa, Wako, Saitama, 351-0198, Japan*

**Abstract**

Recently, ternary clathrate hydrides are promising candidates for high-temperature superconductor. However, it is a formidable challenge to effectively hunt high-temperature superconductivity among multinary hydrides due to the expensive computational cost associated with large unit cells and huge stoichiometric choices. Here we present an efficiently data-driven strategy, including generated clathrate frameworks, the quick estimation of stability for each framework and superconducting critical temperature ($T_c$) for each hydride structure, to accelerate the discovery of high-temperature superconducting hydrides. Our strategy was initialized with more than one million input structures via zeolite databases and our generated dataset. As a result, such a strategy hitherto uncovered 14 prototypical hydrogen frameworks for clathrate hydrides, which is 1.5 times greater than the number (9) of previously reported prototypes. Remarkably, eleven ternary clathrate structures were predicted to have $T_c$s above 250 K at 300 GPa. Further extensive global structure-searching simulations support that $Li_2NaH_{17}$ and $ThY_2H_{24}$ are thermodynamically stable at 220 and 150 GPa, respectively, with $T_c$s approaching room temperature of 297 K and 303 K, which are promising for future synthesis. These results offer a platform to explore high-temperature superconductors via a great number of databases.

**Introduction**

Armed with state-of-the-art crystal structure prediction (CSP) method [1,2], extensive structure-searching simulations establish the first-ever sodalite-like clathrate superhydride (hydrides with unexpectedly high hydrogen contents) $CaH_6$ with a high $T_c$ of 220-235 K at 150 GPa in 2012 [3], in which the peculiar hydrogen cages ($H_{24}$) play a crucial role in determining a large electron-phonon coupling strength and the resultant high superconductivity. This prediction generated a surge of interest to search for over 200 K and even near-room-temperature superconductors among clathrate superhydrides [4,5]. In line with this approach, the later theoretical prediction indeed offers a high-temperature candidate of $LaH_{10}$ [5], which is synthesized by two independent groups and sets the record of superconducting critical temperature (250 K at ~180 GPa) [6,7]. Moreover, this clathrate structure of $CaH_6$ is also successfully synthesized by two independent groups [8,9] recently. Besides, a series of theory-oriented clathrate superhydrides were also successfully synthesized, such as $YH_6$ and $YH_9$ with $T_c$s of 220-243 K at 166-201 GPa [10-12].

Much effort, in addition to binary hydrides, was devoted to the exploration of ternary hydrides due to their more diverse structures and electronic properties [13-16]. On the theoretical hand, a metastable ternary hydride of $Li_2MgH_{16}$ was predicted to have a $T_c$ of 473 K at 250 GPa [17]. Moreover, a great number of ternary hydrides were predicted with $T_c$ near or even above room temperature at megabar pressures [18-24]. On the other hand, several alloy hydrides were identified with high $T_c$ under high pressure, such as $(La, Y)H_{10}$ with a $T_c$ of 253 K at 183 GPa [25] and $(La, Ce)H_9$ [26] with a $T_c$ of 178 K. Very recently, two ternary hydrides of $LaBeH_8$ [27] and $LaB_2H_8$ [28] were synthesized at around 100 GPa.

Although these remarkable findings indicate that clathrate hydrides are promising candidates of high-temperature superconductors, one of the pressing challenges is the search for crystal structure with high superconducting critical temperature effectively due to their much more complex stoichiometry and large unit cells. Besides the CSP methods, several approaches have been developed to identify and predict high-$T_c$ superconductors through high-throughput methods and machine learning [29-31]. To the best of our knowledge, fewer than ten prototypes of clathrate superhydrides have been identified in previous studies [3,6,17,32-37]. It is noted that most prototypes of previously predicted and synthesized clathrate superhydrides can be found in zeolite databases [38,39], which contain a collective of 3D pore framework aluminosilicates composed of $TO_4$ tetrahedra (T is Al or Si). For example, the hydrogen frameworks of $CaH_6$ [3], $LaH_{10}$ [6], $Eu_8H_{46}$ [34], $Li_2MgH_{16}$ [17] could be considered as 'SOD', 'AST', 'MEP' and 'MTN', respectively. It's worth noting that only a small fraction of zeolite frameworks has been identified through the structure prediction of superhydrides. It is of fundamental interest to perform data mining on zeolite databases to explore new potential high-temperature superconducting frameworks.

Toward this goal, here we develop a three-step workflow (see Figure 1) for rapidly screening clathrate frameworks from the databases, which begins with high-throughput calculations of more than 920,000 structures from three zeolite databases: the Database of Zeolite Structure [38] (referred to as the Zeolite), the Reticular Chemistry Structure Resource [39] (RCSR), and a hypothetical database known as the predicted crystallography open database [40] (PCOD).

The workflow is also examined by our generated structure dataset, which contains 300,000 uncategorized structures. As a result, a total of 14 clathrate frameworks are recognized as superhydrides prototypes (6 from the RCSR, 4 from the PCOD and 4 from our generated dataset), where details of frameworks are listed in Table S1. Among them, seven frameworks exhibit high-$T_c$ superconductors above 250 K at 300 GPa ($P6_3/mmc$-$A_2BH_{17}$, $I4/mcm$-$AB_2H_{17}$, $P6/mmm$-$A_5B_2H_{40}$, $I$-$43m$-$A_3B_4H_{41}$, $R$-$3m$-$A_9B_4H_{74}$, $Pnma$-$ABH_{12}$, and $P$-$3m1$-$AB_2H_{21}$. A and B represent metal elements), where the framework structures are shown in Figure 2. Notably, $P6_3/mmc$-$Li_2NaH_{17}$ can be thermodynamically stable at 220 GPa with a $T_c$ of 297 K. Such structure adopted 80 atoms per unit cell, which is difficult to be realized in traditional structure-searching simulations. Besides, our strategy also successfully identified the framework of $P6/mmm$-$AB_2H_{24}$ from our generated structures dataset, which is the prototype of room-temperature superconductor LaSc$_2$H$_{24}$ [36]. *P6/mmm*-ThY$_2$H$_{24}$ is thermodynamically stable around 112 GPa with a $T_c$ of 303 K.

The subsequent sections of this paper are structured as follows. Section II provides the details of the three-step workflow strategy, including generated clathrate frameworks, quick estimation of the stability for a framework as well as superconducting temperatures for each hydride structure. Section III shows the superconductivity of predicted structures from the databases. In Section IV, we conduct a statistical analysis by focusing on how the selection of metal elements tunes the stability and superconductivity of a hydride. Finally, the conclusions are summarized in Section V.

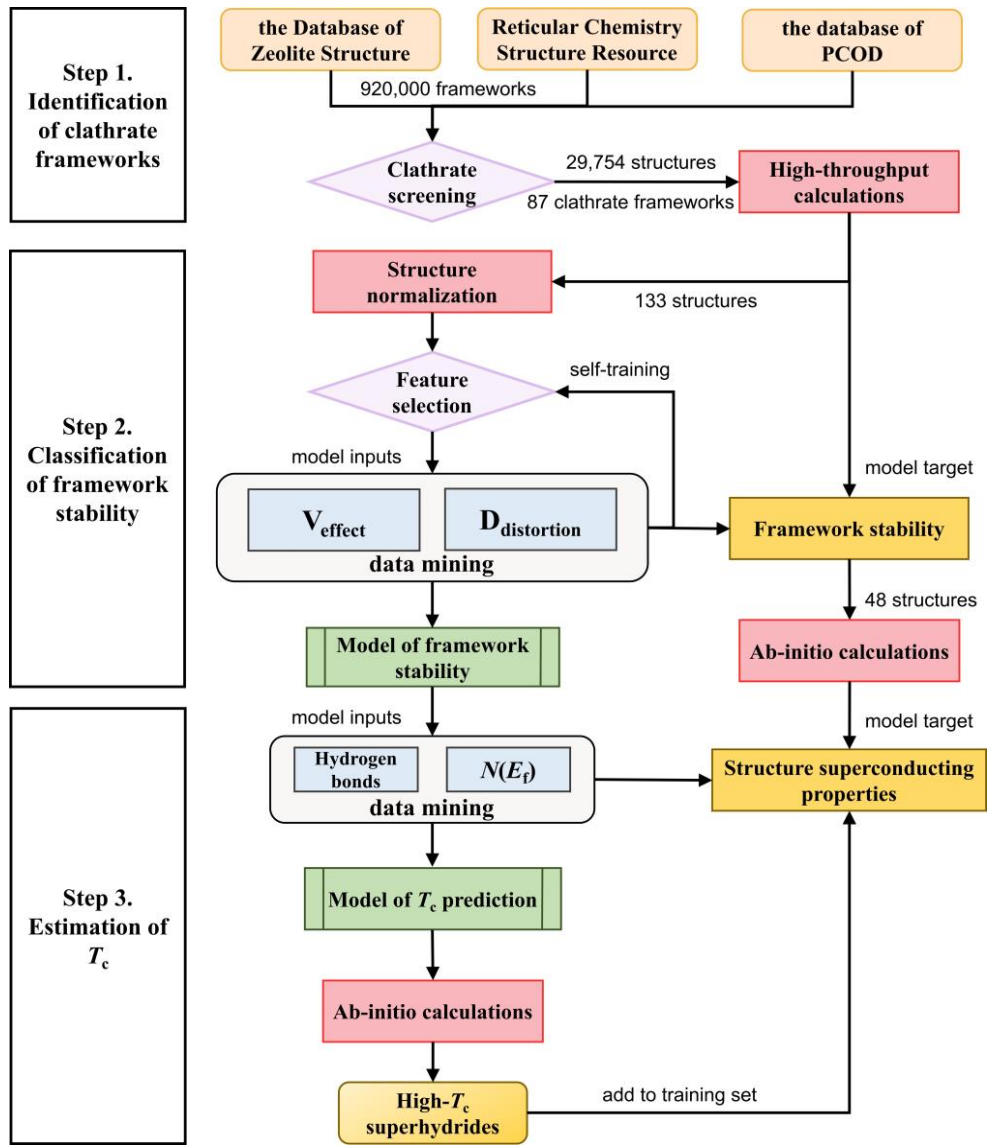

Figure 1. The workflow of our strategy includes three steps: identification of clathrate

frameworks, classification of framework stability and estimation of $T_c$. In step 1, we collected all the frameworks from the zeolite databases, and identified suitable clathrate frameworks by our clathrate screening program. The remaining clathrate frameworks were then inserted by metal elements in the cage units and followed by high-throughput calculations at 300 GPa. In step 2, the optimized structures were categorized by dynamical stability and utilized for the training of classification models. In step 3, we calculated the values of $T_c$ of the dynamically stable structures as model target and collected their electronic and structural properties as input features for the $T_c$ prediction model. Once the workflow was completed, we tested our strategy using structures both from our generated data and previous studies. The results of the workflow can be added to the training set to further refine the classification and estimation models.

**Strategy of Three-step Workflow**

The workflow scheme, as shown in Figure 1, begins with the classified and selected frameworks of the databases before the high-throughput calculations (HTC), where three zeolite databases are collected as follows. The Zeolite [38] and the RCSR [39] are referred to as the experimental dataset, containing over 3,500 naturally discovered zeolite frameworks, irrespective of chemical compositions. The other dataset is the PCOD [40], a hypothetical predicted zeolite database with over 920,000 structures.

**1. Identification of clathrate frameworks**

As frameworks in the Zeolite and RCSR have been categorized based on tiling styles and composite building units in the databases (e.g., chiral, weaving, channel and net), here we select 61 clathrate frameworks that contain at least one cage unit and exclude those incorporate channels and uniformly structured network features. We excluded the frameworks that have been previously discovered (SOD, AST, MEP and MTN). The oxygen atoms are eliminated and vertices are replaced by hydrogen atoms. Furthermore, the frameworks are rescaled to adopt a typical H-H distance of around 1 Å in clathrate hydrides [13] [41].

Since the PCOD database contains hundreds of thousands of unclassified frameworks, it is difficult to manually identify clathrate frameworks. Additionally, during the HTC procedure, a filtering scheme is also necessary for efficient screening clathrate structures. Therefore, we propose a scheme to rapidly identify structures with cage units, which could be employed in the HTC process (as depicted in the rhombic part of Figure 1 Step 1). The scheme is based on references to the cage units that are extracted from clathrate frameworks in the experimental dataset (Fig. S1). It is worth noting that some cage units have been recognized as basic components in the Zeolite database and labeled with three letters (shown in italic font), while others are labeled with point group and number of polygons. These cage units contain many varieties, including most of previously discovered clathrate superhydrides. For instance, $MH_6$ and $MH_{10}$ are composed of cage units sod-24 and ast-32, respectively. $M_8H_{46}$ is composed of cage units mtn-20 and $D_{6d}$-24 (for details, see Table 1). The cage units are normalized with the closest hydrogen-hydrogen distance being around 1 Å. Four criteria for identifying cage units are generalized as follows. First, the cage units should possess moderate vacuum to host metal elements, where the radii parameters range from 1 to 3 Å. Second, the closest distance between two hydrogen atoms should be around 1-1.4 Å. Third, cage units are composed of regular or slightly distorted polygons and the edge of polygons should be fewer than eight. Finally, each hydrogen atom should be connected to at least two other hydrogen atoms, which is a typical feature of clathrate superhydrides. Through our clathrate screening scheme, 26 clathrate frameworks are rapidly identified from 920,000 structures in the PCOD database. Together with the 61 frameworks found in the experimental dataset, we have collected a total of 87 frameworks to generate clathrate superhydride structures.

Subsequently, to produce candidate ternary superhydride structures, we selected two different metal elements, including elements from the first four columns of the periodic table and La, Ce, Ac, Th, to insert into the cage units, resulting in 29,754 initial ternary structures derived from the frameworks. Then we performed geometry-optimization simulations on all structures at 300 GPa. Structures distorted from clathrate features were further screened out after the optimizations. Finally, we found that a total of 133 structures from 27 clathrate frameworks remained. Eleven clathrate frameworks from the experimental dataset and 16 frameworks from the PCOD dataset. We then performed phonon simulations to investigate the dynamical stability of these clathrate structures (see the Methods in the Supplemental Materials).

As a result of the phonon simulations, 6 frameworks from the experimental dataset and 4 frameworks from the PCOD dataset are identified to be dynamically stable at 300 GPa (see Table S1 and Figure 2). In the experimental dataset, the names of frameworks and stoichiometries are as follows: mgz-x-d ($A_2BH_{17}$), cal-d ($AB_2H_{17}$), zra-d ($A_5B_2H_{40}$), mur ($A_9B_4H_{74}$), cbd ($ABH_{11}$) and cla ($A_3B_4H_{41}$). In the PCOD dataset, the indexes of stable frameworks are #8164362 ($A_3B_2H_{28}$), #8252947 ($AB_2H_{12}$), #8101309 ($AB_2H_{22}$) and #8110585 ($AB_2H_{26}$) (A and B represent metal elements).

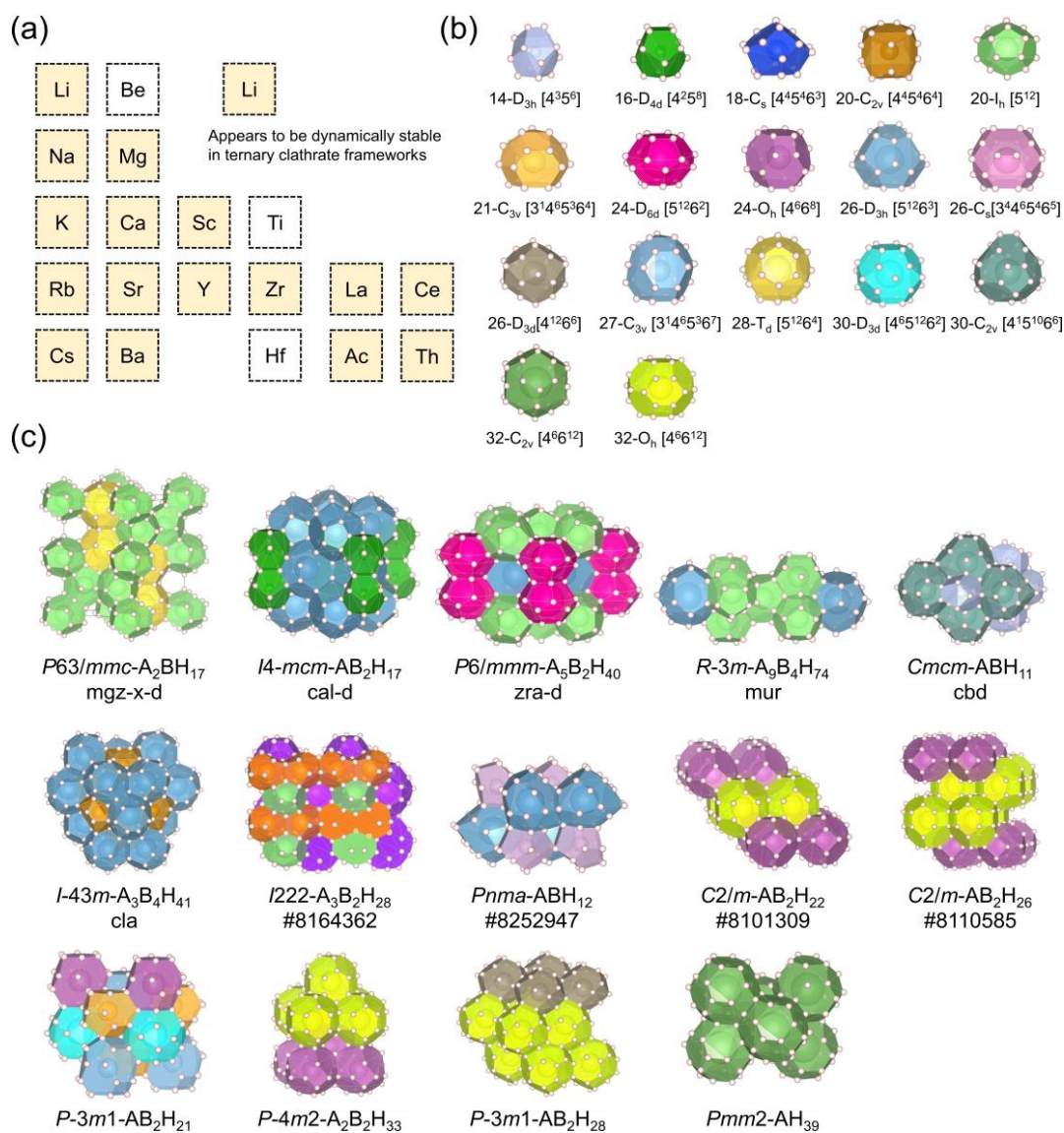

Figure 2. All selected metal elements are listed in (a). Elements from the first four columns of the periodic table, including La, Ce, Ac, and Th were selected for high-throughput calculations. The highlighted elements represent those that were found to form dynamically stable structures during the HTC procedure. The cage units extracted from the dynamically stable frameworks are shown in (b). The vertices of the cage units are represented by H atoms. The cage unit label is named in the form of 'vertices

number-point symmetrical polygons. Fourteen dynamically stable clathrate frameworks are shown in (c). Frameworks in the top two rows are discovered from zeolite databases, along with their labels behind the formula. Frameworks in the third row are predicted from our generated hydrogen dataset. For clarity, the colors of the cage units correspond to those in (b). The frameworks are composed of at least two types of cage units.

**2. Classification of framework stability**

Following the high-throughput calculations, we have also extensively explored candidate structures from the experimental dataset (the Zeolite and RCSR) and the PCOD dataset. As a result, 27 framework prototypes remained clathrate features after the geometry optimization. Further phonon simulations indicate that only ten of them could be dynamically stable, allowing them to serve as promising candidate frameworks, yielding an accuracy of 37%. On the one hand, our newly predicted frameworks enlarge the sample space of clathrate superhydrides. On the other hand, 77 (88%) dynamically unstable clathrate frameworks were indistinguishable from our clathrate screening scheme. The low accuracy of our screening scheme would result in excessive resource consumption in the HTC procedure. This encourages us to train a machine-learning model to more accurately classify the dynamic stability of clathrate frameworks.

The clathrate frameworks of the experimental databases along with their cage units are summarized in Table 1. The first group represents stable frameworks found in previous studies and our newly predicted stable frameworks that exhibit dynamically stable superhydrides. The second group denotes frameworks that retain clathrate features without dynamic stability. As shown in the Table 1, it can be observed that certain cage units, such as $I_h$-$5^{12}$, $D_{6d}$-$5^{12}6^2$ and $T_d$-$5^{12}6^4$ (labeled with point group and number of polygons), appear repeatedly in stable frameworks. In contrast, several cage units only appear repeatedly in unstable frameworks, such as $D_{3h}$-$4^66^5$, $D_{3h}$-$4^66^{11}$ and $D_{6h}$-$5^{12}6^6$. These results indicate that the combination of cage units plays a crucial factor in determining the stability of the framework. Specifically, stable frameworks are composed of 'stable' cage units. With this assumption, the objective of classifying the stability of frameworks is intimately correlated with the classification of suitable cage units.

The dataset for our classification model includes 133 clathrate structures optimized at 300 GPa, involving 376 cage units extracted from these structures. These cage units serve as the basic elements that describe the stability of the frameworks. We labeled those cage units that are extracted from stable frameworks as 'stable' cage units. Considering that frameworks from the experimental databases are more structurally reasonable compared to hypothetical frameworks; to avoid data bias, we particularly accommodate stable frameworks from the experimental databases to the training dataset.

Table 1. The training dataset for the classification model is derived from experimental zeolite frameworks. The cage units of these frameworks are listed along with their point symmetry and the number of polygons. Cage units without an asterisk are in the 'stable region' in Fig. 3 (b) and labeled as 'stable' cage units, while those with an asterisk are labeled as 'unstable' cage units in the dataset.

| Frameworks | Cage units | | |
|---|---|---|---|
| **Dynamically stable** | | | |
| SOD | $O_h$-$4^6 6^8$ | | |
| AST | $O_h$-$4^6 6^{12}$ | | |
| MEP | $I_h$-$5^{12}$ | $D_{6d}$-$5^{12} 6^2$ | |
| MTN | $I_h$-$5^{12}$ | $T_d$-$5^{12} 6^4$ | |
| cal-d | $D_{3h}$-$5^{12} 6^3$ | $D_{4d}$-$4^2 5^8$ | |
| cbd | $D_{3h}$-$4^3 5^6$ | $C_{2v}$-$4^1 5^{10} 6^6$ | |
| cla | $D_{3h}$-$5^{12} 6^3$ | $C_{2v}$-$4^4 5^4 6^4$ | |
| mgz-x-d | $I_h$-$5^{12}$ | $T_d$-$5^{12} 6^4$ | |
| mur | $I_h$-$5^{12}$ | $T_d$-$5^{12} 6^4$ | $D_{6d}$-$5^{12} 6^2$ |
| zra-d | $I_h$-$5^{12}$ | $D_{6d}$-$5^{12} 6^2$ | $D_{3h}$-$5^{12} 6^3$ |
| **Dynamically unstable** | | | |
| alb-x-d | $D_{3h}$-$4^3 5^6$ | *$D_{6h}$-$5^{12} 6^8$ | |
| SGT | $D_{3h}$-$4^3 5^6$ | *$D_{2d}$-$5^{12} 6^8$ | |
| LOS | $D_{3h}$-$4^6 6^{11}$ | *$D_{3h}$-$4^6 6^5$ | |
| FRA | $D_{3h}$-$4^6 6^{11}$ | *$D_{3h}$-$4^6 6^5$ | $O_h$-$4^6 6^8$ |

| DOH | $I_h$-$5^{12}$ | *$D_{6h}$-$5^{12}6^8$ | $D_{3h}$-$4^35^66^3$ |

The scheme of our classification model is shown in Step 2 of Figure 1. Our data mining process underwent a self-training procedure. Each training epoch contains two iterations. In the first iteration, the model focuses on selecting suitable features to classify 'stable' cage units. The score of the model is the classification accuracy of the 'stable' cage units. The classification boundaries were determined by empirical formulas derived from the geometrical features of the cage units. The region of the 'stable' data is marked as the 'stable region'. It is important to note that, at this stage, some unlabeled data were indistinguishable from the 'stable' data. In the second iteration, we use the classification accuracy of the frameworks' stability as a score to refine the boundaries of 'stable region' in reverse and label the remaining unlabeled data based on their positions relative to the 'stable region'. The training process continues until all cage units that are incapable of forming stable clathrate frameworks are excluded from the 'stable region'.

Before training, to mitigate the cage distortions induced by the metal elements, the cage units are pre-normalized according to three standards: 1) the volume of cage units equals one, 2) the radius of sphere equals one, and 3) the average bond length equals one. We utilized various geometric features, such as face density, hydrogen-hydrogen distance, to describe the cage units. Furthermore, we proposed that 'stable' cage units should possess similar characteristics. In the first iteration of training, we used those 'stable' cage units, which appear in stable frameworks, as positive labels to perform self-supervised learning. The formulas of the classification boundaries were provided by the SISSO (Sure Independence Screening and Sparsifying Operator) [42] and the genetic algorithm [43], which can generate high correlation empirical formulas. Once the 'stable' boundaries were established, we were able to identify the dynamic stability of the frameworks. These frameworks can be categorized into four phases. The x and y axes represent, respectively, the summation of the absolute values of boundaries' formulas of their composed cage units and multiplied by a logical 'and' of the 'stable' label ('True' if stay within the 'stable region'). In the second iteration, the classification accuracy of frameworks' stability was used as the target to refine the expressions of boundary formulas reversely. Ultimately, the radii of the cage units were adopted as the best features. Two empirical factors ($D_{distortion}$ and $V_{effect}$) were determined to define the boundaries of the 'stable region'. The results of our classification models for frameworks' stability and 'stable region' of cage units are shown in Figure 3. The empirical expressions are listed as below. The parameters $R_1$, $R_2$, and $R_3$ denote the radii of the three types of normalized cage unit mentioned above, respectively. As depicted in Figure 3 (b), the boundaries of the 'stable region' are $D_{distortion} > 0.54$ and $V_{effect} < 0.12$.

$$D_{distortion} = \frac{R_1 - R_2}{R_1} \quad (1)$$

$$V_{effect} = (R_3 - 1.25)^2 \quad (2)$$

These two formulas describe the geometrical features of the cage units in two dimensions. Equation 1 describes the degree of distortion of a cage unit from a sphere. Since 'stable' cage units prefer a degree of $D_{distortion}$ greater than 0.54, implying $R_1 > 2.17 R_2$, the value of $R_1$ should be as large as possible when the volumes are the similar. Equation 2 represents the effective vacuum space needed by metal atoms, indicating a preferable radius range of 1.25±0.34 Å. Actually, $D_{distortion}$ can be considered as a restriction on the shape of the cage unit itself, while $V_{effect}$ represents the relationship between the volume of cage units and the metal atoms. The results of our classification models reveal that, for a dynamically stable framework, all its constituent cage units remain in the 'stable region'. In contrast, for a dynamically unstable framework, at least one cage unit is found in the 'unstable region', as marked by an asterisk in Table 1.

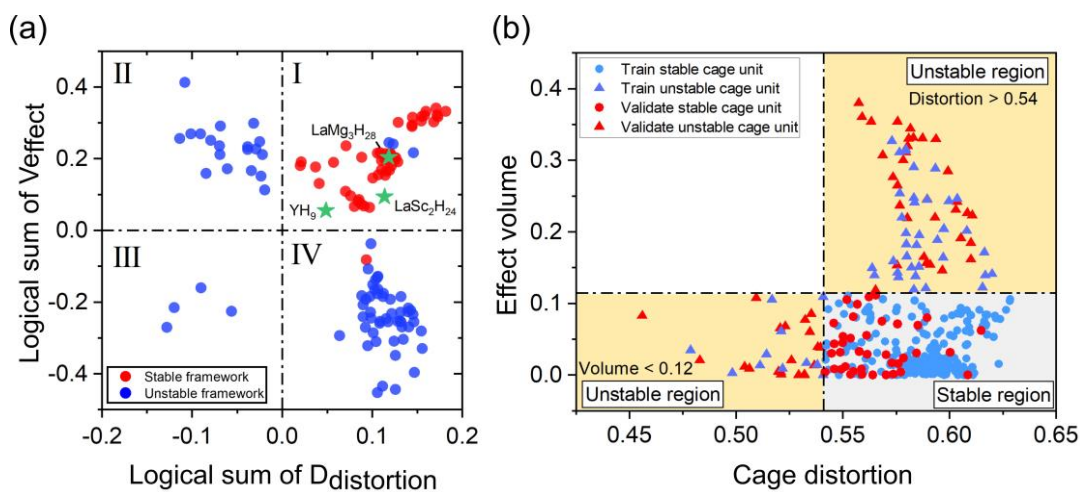

Figure 3. The classification of frameworks' stability in (a) and the classification of stability from the perspective of cage units in (b). The training and validating dataset were collected at the pressure of 300 GPa. In (a), frameworks can be classified into four phases. The x and y axes represent, respectively, the summation of the absolute values of $D_{distortion}$ and $V_{effect}$ of their composed cage units and multiplied by a logical 'and' of the 'stable' label. Phase I represents frameworks composed of 'stable' cage units. Phase II and Phase IV represent frameworks with at least one cage unit that does not satisfy the criteria for cage distortion (Phase II) or effect volume (Phase IV). Phase III indicates that all other criterion is satisfied. Additionally, three non-zeolite structures collected from Refs. [11,35,36] are represented in (a) (indicated by star points). All these structures fall within Phase I, showing good agreement with our classification model. We note that a framework is considered stable only if all its cage units fall within the 'stable region' (i.e., a logical 'and' condition). In (b), based on the degree of cage distortion and the effective volume, the cage units are classified into three regions. Cage units that fall within the 'stable region', where the effective volume < 0.12 and the distortion > 0.54 are labeled as 'table' and represented by red and blue dots.

Finally, we evaluated the accuracy of our classification model. Based on our screening scheme, 27 clathrate frameworks were identified, with 10 frameworks computed to be dynamically stable and 17 frameworks dynamically unstable. The classification model successfully predicted 10 frameworks to be stable and 15 frameworks to be unstable, achieving an accuracy of 92%.

In addition to the frameworks in our datasets, we found that other frameworks were not included in the zeolite databases also fit well with our classification model. We used the $YH_9$ clathrate type [11], recently discovered $LaMg_3H_{28}$ [35] and room-temperature $LaSc_2H_{24}$ [36] as the testing dataset. The cage unit of $YH_9$ is $D_{3h}$-29. The cage units of $LaMg_3H_{28}$ are $D_{6h}$-30 (La) and $D_{2h}$-20 (Mg), and the cage units of $LaSc_2H_{24}$ are $D_{6h}$-30

(La) and $D_{3h}$-24 (Sc). By applying the empirical factors ($D_{distortion}$ and $V_{effect}$) derived from our classification model, these structures are well-suited in the 'stable region', in good agreement with our predictions (as shown in Figure 3 (a)). Taking cage units as input, our model can be implanted in the clathrate screening scheme, providing an efficient approach for exploring dynamically stable clathrate frameworks via databases, particularly in uncategorized databases, such as those generated from CSP and structure generating methods as discussed in Section III.

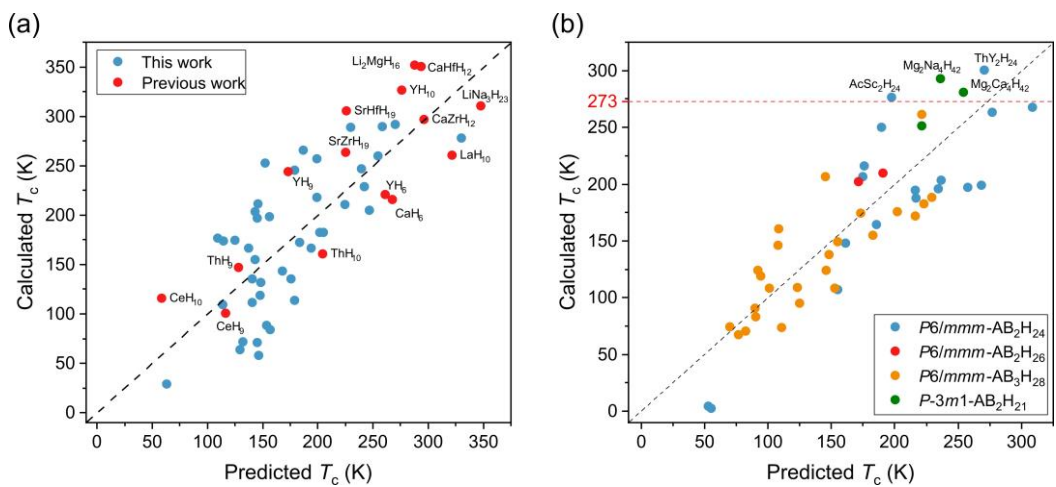

Figure 4. The predicted superconducting temperatures are compared with the results from full electron-phonon simulations based on the Equation (3). Structures from zeolite databases as depicted in (a). The blue dots represent data collected from this article. The red dots are collected from previous studies. Structures from our generated structures and derived from previous discovered studies are plotted in (b). The blue and red dots represent structures that are derived from the metal element substitutions based on the framework of $P6/mmm$-LaSc$_2$H$_{24}$. The yellow dots represent substituted structures from the $P6/mmm$-LaMg$_3$H$_{28}$. The green dots represent structures predicted in our generated dataset. Structures with $T_c$ exceeding 273 K are labeled.

### 3. Quick estimation of clathrate superconducting critical temperature

Following the procedure of HTC calculations, ten zeolite frameworks composed of 40 ternary superhydrides were identified to be dynamically stable at 300 GPa. It is of great interest to estimate their superconducting properties, while full electron-phonon simulations are extremely time-consuming, especially for most of them consist of 40-80 atoms per unit cell. To rapid identify high-$T_c$ superconductors, therefore, here we developed a $T_c$ assessment model to estimate superconductivity for clathrate superhydrides at the end of our workflow, as shown in Step 3 of Figure 1. The training dataset was randomly selected from our identified structures, while the remaining data were used as testing set to further improve the model's performance. Their superconducting critical temperatures ($T_c$) were finally checked by fully electron-phonon simulations using Quantum Espresso (QE) [44].

Previous studies have revealed correlations between electronic properties and critical temperatures in clathrate superhydrides [13,41,45], while the current results not always show that correlation between $T_c$ and electronic properties in our newly predicted structures (Fig. S2). This can be attributed to the diversity of hydrogen-hydrogen distance, which plays a crucial role in determining the strength of electron-phonon coupling (as discussed in Section III). In our estimation model, we consider both hydrogen-hydrogen distance and the projected density of states of hydrogen at the Fermi level ($PDOS_H$). Combining with the SISSO method [42], we introduce an empirical factor called $E_{couple}$ to fit the relationship among $T_c$, electronic properties, and hydrogen-hydrogen distance, as shown in Figure 4 (a). The formula is presented below:

$$E_{couple} = (R_{dosH} * S_{couple}) / (1 + R^2_{dosH}) \tag{3}$$

The $R_{dosH}$ represents the ratio of $PDOS_H$ at the Fermi level. The $S_{couple}$ is defined as the sum of the hydrogen-hydrogen lengths multiplied by the average value of $PDOS_H$ per atom per hydrogen bond. The expression for $S_{couple}$ is given by $\sum_{i=Bl}^{Br} E_H * bond_i$ ($E_H$ in units of states/eV/atom/bond). The $B_l$ and $B_r$ are hyperparameters and represent the range of effective hydrogen-hydrogen lengths, which are determined by the Pearson correlation coefficient score. The optimal values for $B_l$ and $B_r$ are set at 0.98 and 1.28, which correlated with the mid-range of phonon frequencies of hydrogen-hydrogen distances (as discussed in Section III). By performing a linear fitting of $T_c$ and $E_{couple}$, we obtain the form $T_c$=A*$E_{couple}$+B, where A and B represent the slop and intercept value of linear fitting (16370.6 and 24.7 for our dataset), respectively. Finally, the model is validated with previously discovered clathrate structures [3-5,17,33,46-48], as depicted in red dots in Figure 4 (a). The Pearson correlation between the calculated $T_c$ and $E_{couple}$ is 0.83, with a mean absolute error (MAE) of 38 K.

We emphasize that our estimation model offers a preliminary assessment of the $T_c$ of clathrate structures. Our workflow can accelerate the identification of high-$T_c$ superconductors among a large pool of generated structures.

**Results**

In this work, ten clathrate frameworks have been identified from the zeolite databases and 6 frameworks exhibit high-$T_c$ superconductors. More than 40 ternary superhydride superconductors have been generated. Among them, 8 structures were predicted with $T_c$ values exceeding 250 K at 300 GPa (detailed results are summarized in the Table S2.). In this section, we focus on the properties of several typical frameworks that exhibit high $T_c$ values. From the perspective of electron-phonon coupling and the distribution of hydrogen-hydrogen length, our work provides insights into the understanding of mechanism for high-temperature superconductivity.

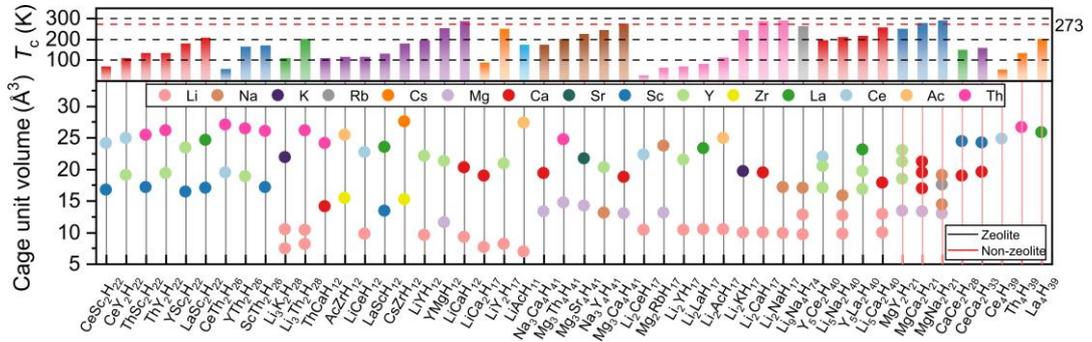

Figure 5. The $T_c$ and elemental distribution of superconductors at 300 GPa. The top panel shows the superconducting temperatures of each compound. 48 structures from 14 frameworks are categorized by different colors. The bottom panel illustrates the combination of metal elements and the volumes of their corresponding cage units within their structures. The black lines represent structures that discovered from the zeolite databases. While the red lines represent structures discovered from our generated structure dataset.

1. Zeolite frameworks

The distribution of structures with respect to $T_c$ is shown in Figure 5. It can be observed that for structures that even adopted the same framework, the combination of metal elements greatly revised the $T_c$ values. In general, combinations such as Li-Na-H, Li-Ca-H, Mg-Na-H and Mg-Ca-H are superconductors with $T_c$ over 250 K. Specifically, $P6_3/mmc$-Li$_2$NaH$_{17}$, $P6_3/mmc$-Li$_2$CaH$_{17}$, $I$-43$m$-Mg$_3$Ca$_4$H$_{41}$ and $Pnma$-LiCaH$_{12}$ could achieve high-$T_c$ values close to room-temperature, reaching 291 K, 289 K, 277 K and 288 K at 300 GPa, respectively.

Furthermore, we found significant $T_c$ differences among the structures in frameworks 'mgz-x-d', 'cal-d' from the RCSR database and framework #8252947 from the PCOD, with a difference of over 160 K between the highest and lowest superconducting temperatures. For example, Li$_2$LaH$_{17}$, Li$_2$YH$_{17}$, LiCa$_2$H$_{17}$ share similar frameworks. However, their $T_c$ values are below 100 K. As listed in Table S2, the $N(E_f)$ ratio of these compounds is relatively large compared to other low $T_c$ superconductors (PDOS$_H$/DOS$_{total}$ > 0.6), exhibiting a divergent trend with $T_c$ values.

The framework of 'mgz-x-d' is composed of T$_d$-28 and I$_h$-20, as shown in Figure 2. In this framework, small metal elements, such as Li and Mg, occupy the smaller cage unit (I$_h$-20), while other metal elements occupy the larger one (T$_d$-28). We found that 'mgz-x-d' shares the same types of cage units as 'MTN' (see Table 1), which is known as type-II clathrate structures (with 40 atoms in unit cell and space group $Fd$-3$m$. Several compounds were found with excellent superconducting properties [47,49,50]. In the case of 'mgz-x-d', the T$_d$-28 cage units are separated in pairs by I$_h$-20. While in 'MTN', the framework contains two sets of networks of T$_d$-28 and I$_h$-20, respectively. It is worth noting that the $Fd$-3$m$-Li$_2$NaH$_{17}$ (MTN) is predicted to be thermodynamical stable at a pressure range of 235-300 GPa in the previous theoretical study [47]. In comparison, the $P6_3/mmc$-Li$_2$NaH$_{17}$ (mgz-x-d) predicted in this work has a larger unit cell of 80 atoms and a lower energy (1.7 meV/atom) than that in energy at 300 GPa.

Among all identified structures, the 'mgz-x-d' framework exhibits the highest $T_c$ values (as listed in Table S2). Fig. S5 shows the calculated electronic band structures and DOS of Li$_2$CaH$_{17}$, Li$_2$KH$_{17}$ and Li$_2$NaH$_{17}$, which exhibit $T_c$ values of 291, 289 and 246 K. The valence electrons of lithium and the other metal element are primarily transferred to hydrogen, as shown in the PDOS calculations with the fact that $N(E_f)$) values for hydrogen reached 1.87, 1.65 and 1.92, respectively. Interestingly, as the pressure decreases from 300 GPa to 220 GPa, the band structure of $Li_2NaH_{17}$ does not change significantly. The value of $N(E_f)$ remains at 1.92 at 220 GPa. However, the electron-phonon coupling constants ($\lambda$) increase from 1.87 to 2.56 as the pressure decreases to 220 GPa, and the $T_c$ increase to 297 K (as shown in Fig. S8). We attribute this enhancement in $\lambda$ to the softening of the frequency of hydrogen-hydrogen interaction. For comparison, we divided the coupling contributions into a set of frequency ranges, as listed in Table S3. It is obvious that in the low-frequency range of around 0-600 $cm^{-1}$, the integration of the coupling increases significantly at 220 GPa, becoming five times larger compared to 300 GPa and contributing 31.8% to the total coupling. This is due to the longer hydrogen-hydrogen distances, as depicted in Fig. S10. At 300 GPa, the hydrogen-hydrogen distance lies in the range of 1.00-1.15 Å.

In contrast, the $T_c$ values of $Li_2AcH_{17}$, $Li_2YH_{17}$, $Li_2LaH_{17}$ and $Mg_2RbH_{17}$ are lower than 100 K. Although the valence electrons of hydrogen still dominate the Fermi level (see the band structures in Fig. S5). However, the value is relatively lower, reaching at most 1.48 in $Li_2AcH_{17}$, compared to those with $T_c$ values exceeding 200 K. Additionally, the larger sizes of the metal atoms enlarge the volume of cage units, as discussed in Section IV, leading to the polarization of the distribution of hydrogen-hydrogen distance (see Fig. S10). The hydrogen-hydrogen distance shifts to the range of 1.10-1.30 and the shortest hydrogen-hydrogen distances are shorter than 1 Å. In the case of $Mg_2RbH_{17}$, the longest distance exceeds 1.55 Å. In Fig. S8, we plot the phonon spectrum of the 'mgz-x-d' superconductors. The main difference between $Li_2CaH_{17}$, $Li_2KH_{17}$, and other low-$T_c$ compounds is the strength of electron-phonon coupling in the range of 1000-2000 $cm^{-1}$, resulting in a final $\lambda$ of only about 0.7 in these compounds.

The 'cal-d' framework is composed of $D_{4d}$-16 and $D_{3h}$-26 units. The unit cell contains 40 atoms and belongs to the space group $I4/mcm$. Two compounds, $LiY_2H_{17}$ and $LiCa_2H_{17}$, are identified as dynamically stable at 300 GPa, with Li occupying $D_{4d}$-16 and Y and Ca occupying $D_{3h}$-26. Despite the same framework, their $T_c$ values differ by more than 160 K. The $LiY_2H_{17}$ has a $T_c$ of 252 K, while $LiCa_2H_{17}$ has a $T_c$ of only 88 K. Interestingly, the difference in $N(E_f)$ values is the opposite (shown in Fig. S6). The $N(E_f)$ of $LiCa_2H_{17}$ is 1.47, 46% greater than that of $LiY_2H_{17}$, which is 1.01. These results defy the common belief that a higher $N(E_f)$ corresponds to a higher $T_c$. From the electron-phonon coupling spectrum (see Fig. S9), the frequency range of $LiCa_2H_{17}$ is higher than that of $LiY_2H_{17}$. For both compounds, the λ comes from the frequency range of 1000-2000 cm$^{-1}$. However, the coupling intensity of $LiY_2H_{17}$ is stronger than that of $LiCa_2H_{17}$, resulting in λ values of 2.0 and 0.79, respectively. The histogram of hydrogen-hydrogen distance is shown in Fig. S11. The hydrogen-hydrogen distances of $LiCa_2H_{17}$ span from 1.0-1.4 Å, with the highest concentration in the range of 1.0-1.1 Å. The hydrogen-hydrogen distances of $LiY_2H_{17}$ are within the range of 1.1-1.2 Å, which contributes to the phonon frequency of 1500-2000 cm$^{-1}$ and enhances the electron-phonon coupling.

We have also performed simulations on electronic and phonon properties of other high-$T_c$ superconductors. We found that in the structures of 'zra-d' and #8164362, which also exhibit high-$T_c$ superconductors above 200 K, the $N(E_f)$ values are generally lower than those in the 'mgz-x-d' and the 'cal-d' frameworks. For instance, (zra-d) $Li_5Na_2H_{40}$ and (#8164362) $Li_3Th_2H_{28}$ exhibit $T_c$ values of 211 K and 203 K, respectively, but their $N(E_f)$ values are only 0.6 and 0.62. Their electron-phonon coupling remains strong in the frequency range of 1000-2000 cm$^{-1}$, reaching λ values of 1.49 and 1.57, respectively (see Fig. S9). Their hydrogen-hydrogen distances are primarily in the range of 1.0-1.2 Å (see Fig. S11).

Finally, these high-temperature superconductivity of our predicted hydrides motivate us to further evaluate their synthesizability. It is known that the synthesizability of a compound is strongly associated with its thermodynamic stability. To verify this, we further performed extensive structure-searching simulations for those systems with high $T_c$ values to identify all possible stable structures. Together with stable structures of previous predicted binary compounds [3,51-58] and structure-searching simulations, we found $P6_3/mmc$-$Li_2CaH_{17}$, $P6_3/mmc$-$Li_2KH_{17}$ and $I4/mmm$-$Sc_2LaH_{22}$ are thermodynamically stable at 300 GPa, with $T_c$s of 289 K, 248 K and 210 K, respectively. Particularly, $P6_3/mmc$-$Li_2NaH_{17}$ can be thermodynamically stable at 220 GPa (see Fig. S3), at which pressure the estimated $T_c$ could approach a room temperature of 297 K.

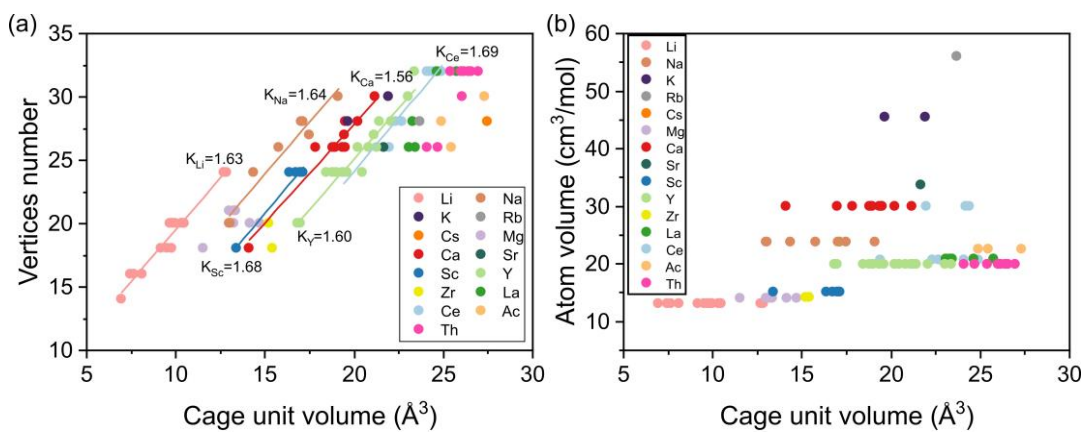

Figure 6. The distribution of cage unit volume with respect to the vertices number (a) and atom volume (b). In (a), the relationship of cage volume and the number of vertices indicates that, in general, the volume of cage unit is proportional to the number of vertices, with a slop around 1.55-1.7. Additionally, the choice of the hosted metal atom indicates the relationship to shift to the right as the radius of the metal atom increases. In (b), the relationship between atom volume and cage volume is depicted. The volume of the cage unit and the metal atom show a step-wise increase for the alkali elements, but there is no correlation between the cage volume and the volume of rare-earth elements.

2. Non-zeolite frameworks

It is worth noting that our strategy can be applied not only to the zeolite databases but also to other existing or generated databases, such as structure generate network [59] and crystal structure prediction methods. To evaluate the capability of our strategy, we generated a large-scale hydrogen dataset from scratch using the Cond-CDVAE approach [59], which can rapidly generate plausible structures using the autoencoder approach. This dataset includes 300,000 structures, each constructed by 30-200 hydrogen atoms and 2-4 randomly selected metal elements in the unit cell. Initially, in the first step of our strategy, the clathrate screening program efficiently filtered out disordered hydrogen frameworks and retained those with cage units. Then, in the second step, the classification model evaluated the stability of these frameworks, predicting four frameworks to be dynamically stable. We then selected elements from the first four columns of the period table to perform high-throughput calculations and estimated their superconducting properties in the third step. Finally, our strategy identified four new types of clathrate frameworks with stoichiometries of $P$-$4m$2-

$A_2B_2H_{33}$, $P$-3$m$1-$AB_2H_{28}$, $Pmm$2-$M_4H_{39}$ and $P$-3$m$1-$AB_2H_{21}$ to be dynamically stable at 300 GPa. The classification model achieved an accuracy of 100%. Among them, $A_2B_2H_{33}$ and $AB_2H_{28}$ are composed of cage units $O_h$-24 and $O_h$-32 and the cage unit of $M_4H_{39}$ is a derivation of $O_h$-32. $AB_2H_{21}$ is composed of four cage units and two of them are new types of cage units ($C_{3v}$-27 and $C_{3v}$-21) as shown in Figure 2. Interestingly, the $T_c$s of $MgCa_2H_{21}$ and $MgNa_2H_{21}$ are 280 and 292 K at 300 GPa, respectively.

Furthermore, our strategy can be applied to predict high-$T_c$ superconductors based on the known clathrate superhydrides. Since clathrate superhydrides are characterized by their embedded metal elements, our strategy could provide the quick estimation of a series of structures with the same framework but various metal elements. By focusing on the structural and electronic properties, our strategy can efficiently identify high-$T_c$ candidates prior to performing full electron-phonon calculations. As a validation, we applied our strategy to hydride frameworks that derived from $P6/mmm$-$LaMg_3H_{28}$ [35] and the room-temperature superconductor $P6/mmm$-$LaSc_2H_{24}$ [36]. We note that the $LaSc_2H_{24}$-type framework was also identified in our generated hydrogen dataset. By substituting metal elements from the first four columns of the periodic table, our strategy predicted six structures with $T_c$s exceeding 250 K. Particularly, we found $P6/mmm$-$ThY_2H_{24}$ reaches 299 K at 200 GPa. As mentioned in Ref. [36], the effects of anharmonic approximation can greatly reduce the pressure required for dynamical stability [60]. Our further structure-searching calculations revealed that $P6/mmm$-$ThY_2H_{24}$ can be thermodynamically stable at pressure below 150 GPa, achieving a $T_c$ of around 290 K by considering the anharmonic effects (see Fig. S4 and Fig. S7).

**Discussion**

In this work, together with the zeolite databases and our generated dataset, we have predicted 14 prototypical clathrate frameworks, 1.5 times more than that in previous studies. The diversity of frameworks and elements enables us to investigate trends in structural and elemental distribution, offering deeper insights into the relationship between structural stability and the choice of metal elements.

From the top panel of Figure 5, it is noticed that most ternary clathrate superhydrides possess high superconducting temperatures, with 40 structures exhibiting $T_c$ values over 100 K and 20 compounds exceeding 200 K. The bottom panel of Figure 5. illustrated the combinations of metal elements with respect to the volumes of their cage units. The distribution results indicate that the volume of the cage unit is related to the type of metal elements rather than the shape of the cage unit itself. Since most clathrate frameworks consist of hydrogen cage units smaller than 15 Å$^3$, lithium becomes the most preferable element.

The volume of the cage units is influenced both by the number of vertices and the metal elements inside, as shown in Figure 6 (a). As the volumes of metal atoms increase, the volume of the cage units also shifts to the right. It is important to note that because of the scarcity of data points, this relationship is only a qualitative trend, and the effects of electronic interactions are not considered. Figure 6 (b) plots the relationship between the volume of the metal atoms and the cage units. The volumes of metal atoms are obtained from the Environmental Chemistry [61] and indicated from the density and mass of bulk crystals at ambient pressure. As the volume of cage units increases, the upper limit of hosted metal volume increases relative to the alkali elements. However, there is no linear correlation between the cage volume and the volume of rare-earth elements. This is likely to be attributed to the fact that the volumes of lanthanide and actinide elements are inferred from bulk materials [61]. Together with the results in Figure 6 (a), the volume of the cage units is associated with the number of vertices and the choice of metal atoms.

**Conclusion**

We have developed a strategy for the large-scale screening of clathrate superconductors. As a result, over 1,220,000 frameworks are collected from the zeolite databases and our generated database. We identified 14 dynamically stable clathrate frameworks. On this basis, our simulations establish 48 superconducting structures, where 20 of them exceeded $T_c$ of 200 K. Remarkably, our additional structure-searching simulations reveal that $P6_3/mmc$-Li$_2$NaH$_{17}$ and $P6/mmm$-ThY$_2$H$_{24}$ are thermodynamically stable at 220 and 150 GPa, respectively, with $T_c$s approaching room temperature of 297 K and 303 K. Our in-depth analysis indicates that, with the cage units as input features, our model to distinguish dynamically stable clathrate frameworks could achieve an accuracy of 92%. The advancement of our current work could be also helpful for the future design of high-temperature superconductors with non-hydrogen clathrate framework, such as the elements contains B, C, N, at near-ambient/ambient pressure [62-64].

## Acknowledgements

This work was supported by the National Key Research and Development Program of China (Grant No. 2022YFA1402304), National Natural Science Foundation of China (Grant No. 52288102, 52090024, 12074138, and 12374007), Program for Jilin University Science and Technology Innovative Research Team (2021TD-05), the Program for Jilin University Computational Interdisciplinary Innovative Platform, the Strategic Priority Research Program of Chinese Academy of Sciences (Grant No. XDB33000000), the Fundamental Research Funds for the Central Universities and computing facilities at the High-Performance Computing Center of Jilin University.

**References:**

[1] Y. Wang, J. Lv, L. Zhu, and Y. Ma, Physical Review B **82**, 094116 (2010).
[2] Y. Wang, J. Lv, L. Zhu, and Y. Ma, Computer Physics Communications **183**, 2063 (2012).
[3] H. Wang, J. S. Tse, K. Tanaka, T. Iitaka, and Y. Ma, Proceedings of the National Academy of Sciences **109**, 6463 (2012).
[4] F. Peng, Y. Sun, C. J. Pickard, R. J. Needs, Q. Wu, and Y. Ma, Physical Review Letters **119**, 107001 (2017).
[5] H. Liu, I. I. Naumov, R. Hoffmann, N. W. Ashcroft, and R. J. Hemley, Proceedings of the National Academy of Sciences **114**, 6990 (2017).
[6] A. P. Drozdov *et al.*, Nature **569**, 528 (2019).
[7] M. Somayazulu, M. Ahart, A. K. Mishra, Z. M. Geballe, M. Baldini, Y. Meng, V. V. Struzhkin, and R. J. Hemley, Phys Rev Lett **122**, 027001 (2019).
[8] L. Ma *et al.*, Physical Review Letters **128**, 167001 (2022).
[9] Z. Li *et al.*, Nature Communications **13**, 2863 (2022).
[10] I. A. Troyan *et al.*, Advanced Materials **33**, 2006832 (2021).
[11] P. Kong *et al.*, Nature Communications **12**, 5075 (2021).
[12] E. Snider, N. Dasenbrock-Gammon, R. McBride, X. Wang, N. Meyers, K. V. Lawler, E. Zurek, A. Salamat, and R. P. Dias, Physical Review Letters **126**, 117003 (2021).
[13] Y. Sun, X. Zhong, H. Liu, and Y. Ma, National Science Review (2023).
[14] X. Zhang, Y. Zhao, and G. Yang, WIREs Computational Molecular Science **12**, e1582 (2022).
[15] B. Lilia *et al.*, Journal of Physics: Condensed Matter **34**, 183002 (2022).
[16] K. P. Hilleke and E. Zurek, Journal of Applied Physics **131**, 070901 (2022).
[17] Y. Sun, J. Lv, Y. Xie, H. Liu, and Y. Ma, Physical Review Letters **123**, 097001 (2019).
[18] Y. Sun, S. Sun, X. Zhong, and H. Liu, Journal of Physics: Condensed Matter **34**, 505404 (2022).
[19] S. Di Cataldo, W. von der Linden, and L. Boeri, npj Computational Materials **8**, 2 (2022).
[20] X. Liang *et al.*, Physical Review B **104**, 134501 (2021).
[21] M. Gao, X.-W. Yan, Z.-Y. Lu, and T. Xiang, Physical Review B **104**, L100504 (2021).
[22] S. Di Cataldo, C. Heil, W. von der Linden, and L. Boeri, Physical Review B **104**, L020511 (2021).
[23] W. Sukmas, P. Tsuppayakorn-aek, U. Pinsook, and T. Bovornratanaraks, Journal of Alloys and Compounds **849**, 156434 (2020).
[24] H. Xie *et al.*, Journal of Physics: Condensed Matter **31**, 245404 (2019).
[25] D. V. Semenok *et al.*, Materials Today **48**, 18 (2021).
[26] J. Bi *et al.*, Nature Communications **13**, 5952 (2022).
[27] Y. Song, J. Bi, Y. Nakamoto, K. Shimizu, H. Liu, B. Zou, G. Liu, H. Wang, and Y. Ma, Physical Review Letters **130**, 266001 (2023).
[28] X. Song *et al.*, Journal of the American Chemical Society **146**, 13797 (2024).
[29] T. F. T. Cerqueira, A. Sanna, and M. A. L. Marques, Advanced Materials **36**, 2307085 (2024).
[30] K. Choudhary and K. Garrity, npj Computational Materials **8**, 244 (2022).
[31] S. Saha, S. Di Cataldo, F. Giannessi, A. Cucciari, W. von der Linden, and L. Boeri, Physical Review Materials **7**, 054806 (2023).
[32] X. Zhong, Y. Sun, T. Iitaka, M. Xu, H. Liu, R. J. Hemley, C. Chen, and Y. Ma, Journal of the American Chemical Society **144**, 13394 (2022).
[33] X. Li *et al.*, Nature Communications **10**, 3461 (2019).
[34] D. V. Semenok *et al.*, The Journal of Physical Chemistry Letters **12**, 32 (2021).

[35] G. M. Shutov, D. V. Semenok, I. A. Kruglov, and A. R. Oganov, Materials Today Physics **40**, 101300 (2024).
[36] X. He, W. Zhao, Y. Xie, A. Hermann, R. J. Hemley, H. Liu, and Y. Ma, Proceedings of the National Academy of Sciences **121**, e2401840121 (2024).
[37] W. Zhao, H. Song, M. Du, Q. Jiang, T. Ma, M. Xu, D. Duan, and T. Cui, Phys. Chem. Chem. Phys. **25**, 5237 (2023).
[38] Database of Zeolite Structures, http://www.iza-structure.org/databases/.
[39] Reticular Chemistry Structure Resource, http://rcsr.anu.edu.au/.
[40] PCOD Database, https://mwdeem.org/PCOD/.
[41] F. Belli, T. Novoa, J. Contreras-García, and I. Errea, Nature Communications **12**, 5381 (2021).
[42] R. Ouyang, S. Curtarolo, E. Ahmetcik, M. Scheffler, and L. M. Ghiringhelli, Physical Review Materials **2**, 083802 (2018).
[43] gplearn, https://gplearn.readthedocs.io/en/stable/intro.html.
[44] P. Giannozzi *et al.*, Journal of Physics: Condensed Matter **21**, 395502 (2009).
[45] T. Ma, Z. Zhang, M. Du, Z. Huo, W. Chen, F. Tian, D. Duan, and T. Cui, Materials Today Physics **38**, 101233 (2023).
[46] L. Liu *et al.*, Physical Review B **107**, L020504 (2023).
[47] D. An, D. Duan, Z. Zhang, Q. Jiang, H. Song, and T. Cui, arXiv e-prints, arXiv: 2303.09805 (2023).
[48] D. V. Semenok *et al.*, Materials Today **33**, 36 (2020).
[49] Y. Sun, Y. Wang, X. Zhong, Y. Xie, and H. Liu, Physical Review B **106**, 024519 (2022).
[50] M. Redington and E. Zurek, Chemistry of Materials **36**, 8412 (2024).
[51] J. Hooper and E. Zurek, The Journal of Physical Chemistry C **116**, 13322 (2012).
[52] D. V. Semenok, I. A. Kruglov, I. A. Savkin, A. G. Kvashnin, and A. R. Oganov, Current Opinion in Solid State and Materials Science **24**, 100808 (2020).
[53] D. Zhou, X. Jin, X. Meng, G. Bao, Y. Ma, B. Liu, and T. Cui, Physical Review B **86**, 014118 (2012).
[54] E. Zurek, R. Hoffmann, N. W. Ashcroft, A. R. Oganov, and A. O. Lyakhov, Proceedings of the National Academy of Sciences **106**, 17640 (2009).
[55] D. C. Lonie, J. Hooper, B. Altintas, and E. Zurek, Physical Review B **87**, 054107 (2013).
[56] P. Baettig and E. Zurek, Physical Review Letters **106**, 237002 (2011).
[57] A. M. Shipley, M. J. Hutcheon, R. J. Needs, and C. J. Pickard, Physical Review B **104**, 054501 (2021).
[58] H. Liu, I. I. Naumov, R. Hoffmann, N. W. Ashcroft, and R. J. Hemley, Proceedings of the National Academy of Sciences **114**, 6990 (2017).
[59] X. Luo, Z. Wang, P. Gao, J. Lv, Y. Wang, C. Chen, and Y. Ma, 2024), p. arXiv:2403.10846.
[60] L. Monacelli, R. Bianco, M. Cherubini, M. Calandra, I. Errea, and F. Mauri, Journal of Physics: Condensed Matter **33**, 363001 (2021).
[61] EnvironmentalChemistry, https://environmentalchemistry.com/yogi/periodic/.
[62] L. Zhu *et al.*, Phys. Rev. Res. **5**, 013012 (2023).
[63] N. Geng, K. P. Hilleke, L. Zhu, X. Wang, T. A. Strobel, and E. Zurek, Journal of the American Chemical Society **145**, 1696 (2023).
[64] L. Zhu *et al.*, Science Advances **6**, eaay8361 (2020).